\documentclass{aa}				

\usepackage{times}				
\usepackage{graphics}

\thesaurus{13         
           <(08.19.5;  
            11.09.1;  
            11.16.1;  
            11.16.6;  
            13.07.1)} 

\title{The host galaxy and optical light curve of the gamma-ray burst
       GRB~980703\thanks{Based on observations made with the NASA/ESA
       \emph{Hubble Space Telescope} (\emph{HST}), obtained at the
       Space Telescope Science Institute, which is operated by the
       Association of Universities for Research in Astronomy, Inc.,
       under NASA contract NAS5-26555.}\fnmsep\thanks{Based on
       observations made with the Nordic Optical Telescope (NOT),
       operated on the island of La Palma jointly by Denmark, Finland,
       Iceland, Norway, and Sweden.}\fnmsep\thanks{Based on
       observations made at the 2.2-m telescope of the German-Spanish
       Calar Alto Observatory}}

\author{S. Holland \inst{1,2}
     \and
        J. P. U. Fynbo \inst{3}
     \and
        J. Hjorth \inst{4}
     \and
        J. Gorosabel \inst{5}
     \and
        H. Pedersen \inst{4}
     \and
        M. I. Andersen \inst{6}
     \and
        A. Dar \inst{7}
     \and
        B. Thomsen \inst{1}
     \and
        P. M{\o}ller \inst{3}
     \and
        G. Bj{\"o}rnsson \inst{8}
     \and
        A. O. Jaunsen \inst{9}
     \and
        P. Natarajan \inst{10}
     \and
        N. Tanvir \inst{11}
}

\institute{Institute of Physics and Astronomy,
           University of Aarhus,
           DK--8000 {\AA}rhus C.,
           Denmark
           e-mail: bt@ifa.au.dk
     \and
           Department of Physics,
           University of Notre Dame,
           Notre Dame, IN 46556--5670,
	   U.S.A.,
           e-mail: sholland@nd.edu
     \and
           European Southern Observatory,
	   Karl-Schwarzschild-Stra{\ss}e 2,
           D--85748 Garching,
           Germany,
	   email: jfynbo@eso.org, pmoller@eso.org
     \and
           Astronomical Observatory,
           University of Copenhagen,
           Juliane Maries Vej 30,
           DK--2100 Copenhagen {\O},
           Denmark,
           email: jens@astro.ku.dk, holger@astro.ku.dk
     \and
           Danish Space Research Institute,
           Juliane Maries Vej 30,
           DK--2100 Copenhagen {\O},
           Denmark,
           email: jgu@dsri.dk
     \and
           Division of Astronomy,
           University of Oulu,
           P. O. Box 3000,
           FIN--90014 Oulu,
           Finland,
           email: manderse@sun3.oulu.fi
     \and
           Department of Physics,
	   Technion, Haifa 32000,
	   Israel,
           email: arnon.dar@cern.ch
     \and
           Science Institute,
           Dunhagi 3,
           University of Iceland,
           IS--107 Reykjavik, Iceland,
           e-mail: gulli@raunvis.hi.is
     \and
           European Southern Observatory,
	   Casilla 19001,
           Santiago 19,
           Chile,
           ajaunsen@eso.org
     \and
           Department of Astronomy,
           Yale University,
           New Haven, CT 06520--8181,
           U.S.A.,
           e-mail: priya@astro.yale.edu
     \and
           Department of Physical Sciences,
           University of Hertfordshire,
           College Lane, 
           Hatfield, Hertfordshire, AL10 9AB,
           U.K.,
           email: nrt@star.herts.ac.uk    
}

\offprints{S. Holland}
\mail{sholland@nd.edu}

\date{Received  / Accepted }

\begin{document}

\titlerunning{The host galaxy of GRB~980703}
\authorrunning{Holland, et~al.}

\maketitle


\begin{abstract}

     We present deep \emph{HST}/STIS and ground-based photometry of
the host galaxy of the gamma-ray burst \object{GRB~980703} taken 17,
551, 710, and 716 days after the burst.  We find that the host is a
blue, slightly over-luminous galaxy with $V_\mathrm{gal} = 23.00 \pm
0.10$, ${(V\!-\!R)}_\mathrm{gal} = 0.43 \pm 0.13$, and a centre that
is $\approx 0.2$ mag bluer than the outer regions of the galaxy.  The
galaxy has a star-formation rate of $8$--$13$ $\mathcal{M}_{\sun}
\mathrm{yr}^{-1}$, assuming no extinction in the host.  We find that
the galaxy is best fit by a Sersic $R^{1/n}$ profile with $n \approx
1.0$ and a half-light radius of $0\farcs13$ ($= 0.72 h_{100}^{-1}$
proper kpc).  This corresponds to an exponential disk with a scale
radius of $0\farcs22$ ($= 1.21 h_{100}^{-1}$ proper kpc).  Subtracting
a fit with elliptical isophotes leaves large residuals, which suggests
that the host galaxy has a somewhat irregular morphology, but we are
unable to connect the location of \object{GRB~980703} with any special
features in the host. The host galaxy appears to be a typical example
of a compact star forming galaxy similar to those found in the Hubble
Deep Field North.  The $R$-band light curve of the optical afterglow
associated with this gamma-ray burst is consistent with a single
power-law decay having a slope of $\alpha = -1.37 \pm 0.14$.  Due to
the bright underlying host galaxy the late time properties of the
light-curve are very poorly constrained.  The decay of the optical
light curve is consistent with a contribution from an underlying Type
Ic supernova like \object{SN1998bw}, or a dust echo, but such
contributions cannot be securely established.


\end{abstract}


\section{Introduction\label{SECTION:intro}}

     \object{GRB~980703} (BATSE trigger 6891) was detected on 1998
July 3.182 UT by the All-Sky Monitor on the \emph{Rossi $X$-Ray Timing
Explorer} satellite (Levine et~al.~\cite{LMM1998}).  Frail
et~al.~(\cite{FHB1998}) and Zapatero~Osorio et~al.~(\cite{ZOC1998})
identified a fading optical afterglow (OA), and Djorgovski
et~al.~(\cite{DKG1998}) derived a redshift of $z = 0.9661 \pm 0.0001$
from absorption and emission lines in the combined spectrum of the OA
and host galaxy.  They also derived a star-formation rate of $\approx
10 \mathcal{M}_{\sun} \mathrm{yr}^{-1}$ for the host galaxy from the
strength of the [\ion{O}{ii}] emission line.  They stressed that the
star formation rate depends on the adopted extinction and on the
star-formation indicator used, but concluded that the host is a
compact starburst galaxy.  $R$-band optical light curves of the OA $+$
host galaxy were presented by Bloom et~al.~(\cite{BFK1998}),
Castro-Tirado et~al.~(\cite{CTZ1999}), and Vreeswijk
et~al.~(\cite{VGO1999}), who all found that the optical light decayed
as a single power law with a slope of $\alpha \approx -1.4$, and all
estimated the magnitude of the host galaxy to be $R_\mathrm{gal}
\approx 22.6$.  Sokolov et~al.~(\cite{SFK2001}) used late-time
photometry of the OA $+$ host to estimate global star-formation rates
in the host galaxy of $15$--$66 \mathcal{M}_{\sun} \mathrm{yr}^{-1}$,
depending on the details of the initial mass function.

     The host galaxy of \object{GRB~980703} appears to have a high
extinction along the line of sight to the gamma-ray burst (GRB).
Castro-Tirado et~al.~(\cite{CTZ1999}) found $A_V = 2.2$ from the slope
of $X$-ray/optical/infrared fluxes 0.9 days after the burst, Vreeswijk
et~al.~(\cite{VGO1999}) used a similar method to find $A_V = 1.50 \pm
0.11$ 1.2 days after the burst, Djorgovski et~al.~(\cite{DKB1998})
found $A_V \approx 0.3 \pm 0.3$ from the Balmer decrement 4.4 days
after the burst, and Bloom et~al.~(\cite{BFK1998}) estimated $A_V =
0.9 \pm 0.2$ 5.3 days after the burst.  The different (and formally
inconsistent) values of $A_V$ may be due to different methods for
measuring the extinction probing different regions of the host galaxy,
or may indicate that the environment where the burst occurred is
changing with time.

     We have obtained new ground-based and \emph{Hubble Space
Telescope} (\emph{HST}) Space Telescope Imaging Spectrograph (STIS)
images of the host galaxy of \object{GRB~980703} taken 17, 577, 710,
and 716 days after the burst.  These data are discussed in
Sect.~\ref{SECTION:obs}.  The nature of the host galaxy and the
location of the GRB within the galaxy are discussed in
Sect.~\ref{SECTION:host}, and the light curve of the OA is discussed
in Sect.~\ref{SECTION:light}.  Unless otherwise noted we have assumed
a cosmology where $H_0 = 100h_{100}$ km s$^{-1}$ Mpc$^{-1}$, $\Omega_m
= 0.3$, and $\Omega_\Lambda = 0.7$.  For this cosmology a redshift of
0.9661 corresponds to a luminosity distance of $4.43h_{100}^{-1}$ Gpc
and a distance modulus of $43.23 - 5\log_{10}(h_{100})$.  One
arcsecond corresponds to $10.9 h_{100}^{-1}$ comoving kpc or $5.6
h_{100}^{-1}$ proper kpc, and the lookback time is $5.3h_{100}^{-1}$
Gyr.


\section{Observations and Data Reductions\label{SECTION:obs}}

\subsection{Ground-based observations\label{SECTION:obs_ground}}

     We obtained $R$- and $I$-band images of the field containing
\object{GRB~980703} using the Andaluc{\'\i}a Faint Object Spectrograph
and Camera (ALFOSC) on the 2.56-meter Nordic Optical Telescope (NOT)
at La Palma on 1998 July 20.172 (16.99 days after the burst). A log of
all our observations is given in Table~\ref{TABLE:obs}.  The
instrumental gain was $1.0$ e$^{-1}$/ADU and the read-out noise was
$6$ e$^{-1}$/pixel.  We calibrated the $R$-band data using six stars
from Rhoads et~al.~(\cite{RDC1998}), and the $I$-band data using three
stars from Vreeswijk et~al.~(\cite{VGO1999}). The magnitudes were
measured in apertures with radii of 15 pixels ($= 2\farcs82$).  The
derived relations between the instrumental and standard systems are $R
= r + (8.440 \pm 0.011) + (-0.074 \pm 0.023)(r\!-\!i)$ and $I = i +
(7.934 \pm 0.038)$. The RMS of the residuals are 0.07 and 0.09 mag
respectively. We found combined host$+$OA magnitudes of $R = 22.40 \pm
0.07$ and $I = 21.71 \pm 0.09$ in an aperture of radius $1\farcs524$
($= 8.5 h_{100}^{-1}$ proper kpc) centred on the OA\@.  The full-width
at half-maxima (FWHMs) of the stellar profiles on the combined NOT
image were $0\farcs8$ in the $R$ band, and $0\farcs9$ in the $I$ band.

     In order to determine the magnitude of the host galaxy without
contamination from the OA we obtained $V$- and $R$-band images of the
host galaxy with the Calar Alto Faint Object Spectrograph (CAFOS) at
the Centro Astron{\'o}mico Hispano-Alem{\'a}n (CAHA) Observatory's
2.2-meter telescope on 2000 January 5 (551 days after the burst).
These data have a gain of $2.3$ e$^{-1}$/ADU, a read-out noise of
$5.06$ e$^{-1}$/pixel, and were calibrated using the same field stars
and apertures as for the NOT data.  The derived relations between the
instrumental and standard systems are $V = v + (5.943 \pm 0.051) +
(0.079 \pm 0.072)(v\!-\!r)$ and $R = r + (6.145
\pm 0.021) + (0.245 \pm 0.030)(v\!-\!r)$.  The RMS of the residuals in
the calibrations are 0.031 and 0.027 mag in $V$ and $R$
respectively.  For the host galaxy we measured $V_\mathrm{gal} = 23.00
\pm 0.10$ and $R_\mathrm{gal} = 22.57 \pm 0.09$ in an aperture of
radius $1\farcs524$.  The CAHA combined $V$-band image had FWHM $=
1\farcs5$ while the combined $R$-band image had FWHM $= 1\farcs7$.

%
\begin{table}
\begin{center} 
\caption{The log of the observations.}
\smallskip
\begin{tabular}{lcll}
\hline
\hline
Telescope/Instrument & Filter & Date (UT) & Exposure Time (s) \\
\hline
NOT/ALFOSC           & $R$ & 1998 June 20  & $5 \times ~~600$ \\
                     & $I$ &               & $5 \times ~~600$ \\
CAHA/CAFOS           & $V$ & 2000 Jan.\ 5  & $1 \times ~~980$ \\
                     &     &               & $1 \times 1200$ \\
                     & $R$ &               & $3 \times ~~900$ \\
\emph{HST}/STIS      & LP  & 2000 June 12  & $4 \times ~~650$ \\
                     &     &               & $4 \times ~~666$ \\
                     & CL  & 2000 June 18  & $1 \times ~~~~60$ \\
                     &     &               & $1 \times ~~615$ \\
                     &     &               & $3 \times ~~613$ \\
                     &     &               & $4 \times ~~666$ \\
\hline
\hline
\end{tabular}
\label{TABLE:obs}
\end{center}
\end{table}
%

     \object{GRB~980703} is located at $\alpha =$ 23:59:06.72 $\delta
= +$08:35:07.3 (J2000) which correspond to Galactic coordinates of
$\ell^\mathrm{II} = 101\fdg56$, $b^\mathrm{II} = -52\fdg10$, hence the
Galactic extinction is small.  We have adopted a Galactic reddening of
$E_{B\!-\!V} = 0.058 \pm 0.020$ (corresponding to extinctions of $A_V
= 0.19 \pm 0.07$, $A_R = 0.16 \pm 0.05$, and $A_I = 0.11 \pm 0.04$)
from the DIRBE/IRAF dust maps of Schlegel et~al.~(\cite{SFD1998}). The
dereddened magnitudes of the host are $V_{0,\mathrm{gal}} = 22.81 \pm
0.12$ and $R_{0,\mathrm{gal}} = 22.41 \pm 0.10$.

     If we assume that the host galaxy has a power law spectrum of the
form $f \propto \nu^\beta$ then we derive $\beta_\mathrm{gal} = -1.08
\pm 0.59$, which gives $I_{0,\mathrm{gal}} = 21.91 \pm 0.23$.

\subsection{The HST data\label{SECTION:hst_obs}}

     We used the \emph{HST}/STIS to obtain CCD images of the host
galaxy of \object{GRB~980703} as part of the Cycle 9 proposal GO-8640.
The total exposure times were 5178 seconds in the 50CCD (clear;
hereafter referred to as CL) aperture and 5264 seconds in the F28X50LP
(long pass; hereafter referred to as LP) aperture.  The CL images were
taken on 2000 June 12 (710 days after the burst) and the LP images
were taken on 2000 June 18 (716 days after the burst).  The CCD gain
was set to $1$ e$^-$/ADU and the read-out noise was $4.46$
e$^-$/pixel.  We used a four-point STIS-CCD-BOX dithering pattern with
shifts of 2.5 pixels ($= 0\farcs127$) between exposures.  The data was
pre-processed through the standard STIS pipeline and combined using
the {\sc dither} (v1.2) software (Fruchter \& Hook~\cite{FH2001}) as
implemented in IRAF\footnote{Image Reduction and Analysis Facility
(IRAF), a software system distributed by the National Optical
Astronomy Observatories (NOAO).}(v2.11.3)/STSDAS(v2.1.1).  The
``pixfrac'' parameter was set to $0.6$, and we selected a final output
scale of $0\farcs0254$/pixel.  These observations are part of the
Survey of the Host Galaxies of GRBs ({\small {\tt
http://www.ifa.au.dk/{\~{}}hst/grb\_hosts/index.html}}, Holland
et~al.~\cite{HFT2000a}).

     We measured the total AB magnitude of the host galaxy on the CL
and LP images in an aperture of radius $1\farcs524$.  The zero points
of Gardner et~al.~(\cite{GBB2000}) yield AB magnitudes of $22.87 \pm
0.02$ in the CL aperture and $22.50 \pm 0.03$ in the LP aperture.

\begin{figure*}
\resizebox{\hsize}{!}{\includegraphics{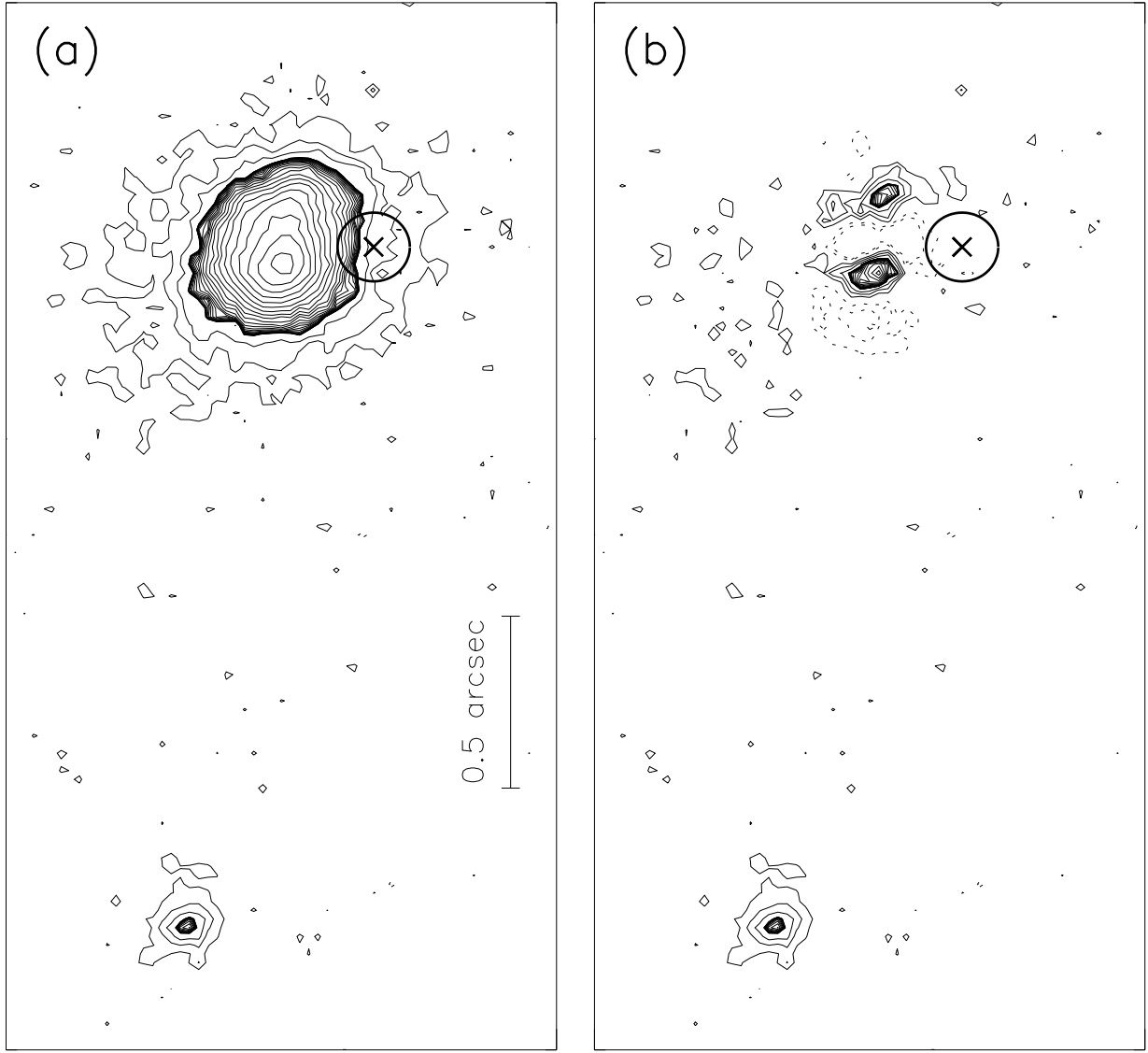}}
\caption{\emph{Left panel}: This Figure shows a $1\farcs6 \times
3\farcs1$ section of the drizzled CL \emph{HST}/STIS image.  North is
up and east is to the left. The lowest contours are shown linearly
whereas the inner contours are logarithmic in order to better show the
properties of the host over a large dynamic range. The ``X'' marks the
location of \object{GRB~980713} and the circle is our 1$\sigma$ error
circle for the astrometric solution.  The possible companion galaxy is
located near the southern edge of the image.  \emph{Right panel}: This
Figure shows the host galaxy with the best-fitting $R^{1/n}$ ($n =
1.07$) model subtracted.  The contour intervals are the same as for
Fig.~\ref{FIGURE:host}a.  The systematic residuals seen near the
centre of the host suggest that there is small-scale structure within
the half-light radius.  There is also an excess of light on the west
side of the host.}
\label{FIGURE:host}
\end{figure*}


\section{The Host Galaxy\label{SECTION:host}}

\subsection{The morphology of the host galaxy\label{SECTION:morphology}}

\begin{figure}
\resizebox{\hsize}{!}{\includegraphics{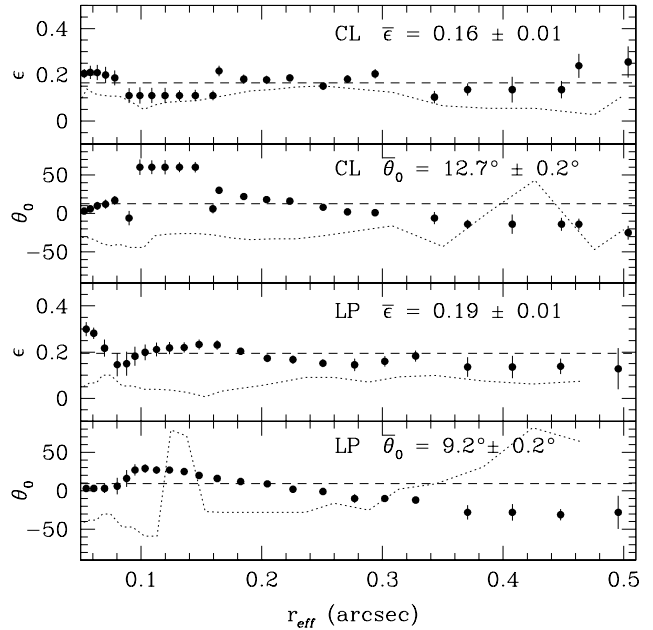}}
\caption{The upper two panels show the observed ellipticity and
orientation profiles for the host galaxy of \object{GRB~980703} in the
CL aperture (filled circles).  The dashed line is the weighted mean of
the ellipticity (or position angle) and the dotted line is the profile
of the PSF\@.  The data points near $0\farcs1$ are constant because
the fitting algorithm was unable to converge at these radii.  The
derived position angles are not reliable for $\epsilon \approx 0$.
The lower two panels show the same thing in the LP aperture.  The data
has not been corrected for the effects of the PSF.}
\label{FIGURE:shape}
\end{figure}

     Fig.~\ref{FIGURE:host}a shows a contour plot of a section of the
STIS CL image containing the host galaxy.  The galaxy is extremely
compact and has a somewhat irregular, ``egg-shaped'' morphology. In
order to quantitatively characterize the morphology of the galaxy we
fitted elliptical isophotes to the CL and LP images of the host using
Jedrzejewski's~(\cite{J1987}) algorithm as implemented in the STSDAS
task {\sc isophote.ellipse}. These isophotes provided a resonable fit
to the galaxy, but there are significant residuals as the isophotes
are not elliptical. The mean observed ellipticity of the host on the
CL image is $\epsilon = 0.16 \pm 0.01$ for $0\farcs05 \le r \le
0\farcs51$, and the mean position angle of the semi-major axis on the
CL image is $\theta_0 = 12\fdg7 \pm 0\fdg2$ east of north on the sky.
We find $\epsilon = 0.19 \pm 0.01$ and $\theta_0 = 9\fdg2 \pm 0\fdg2$
on the LP image.  The variation of the ellipticity and position angle
with radius is shown in Fig.~\ref{FIGURE:shape}.  The host galaxy
shows some evidence for twisted isophotes as the position angle of the
major axis rotates from $\approx +45\degr$ to $\approx -45\degr$
between $0\farcs1$ and $0\farcs5$ from the centre of the host galaxy.
This change is seen in both the CL and LP images. These values have
not been corrected for the effects of the point-spread function (PSF)
on the the observed shape of the galaxy.  Schweizer~(\cite{S1979}) and
Holland~(\cite{H98}) have shown that the observed shape and surface
brightness profile of a galaxy can be dominated by the effects of the
PSF if the diameter of the galaxy is less than $\approx 8$ times the
width of the PSF\@.  The FWHM of the STIS PSF in the CL and LP images
is $\approx 0\farcs084$, so the shape of the PSF dominates the central
regions of the host galaxy, and will effect the observed shape in the
outer regions.  The PSF has a mean ellipticity of $0.11$ and a mean
position angle of $-31\fdg5$ in the CL image.  This is $44\degr$ from
the observed major axis of the host galaxy, so the PSF will act to
make the host appear to be less elliptical than it actually is.

     In order to determine the intrinsic (i.e., before the light
passed through the \emph{HST}/STIS optics) surface-brightness
distribution of the host galaxy we fit two-dimensional PSF-convolved
generalized de Vaucouleurs profiles (Sersic~\cite{S1968})

\begin{equation}
S(R) = S_e 10^{-b(n)/\ln(10) \left[ {(R/R_e)}^{1/n} - 1 \right]}
\label{EQUATION:gdv}
\end{equation}

\noindent
where $R_e$ is the radius enclosing half the light from the galaxy and
$S_e$ is the surface brightness at $R_e$, to the CL and LP images of
the host galaxy using the software described in Holland
et~al.~(\cite{HCH1999}).  This software allows all of the model
parameters to be fit simultaneously using a chi-square minimization
scheme.  The parameters of the best fitting models, along with the
$\chi^2$ per degree of freedom (DOF) are listed in
Table~\ref{TABLE:models}. The number of degrees of freedom in each fit
is the number of pixels used in the fit minus the number of model
parameters.  The constant $b(n)$ is chosen so that $R_e$ encloses half
of the light of the galaxy.  The mean surface brightness inside $R_e$
is denoted by ${\langle\mu\rangle}_e$.  The formal 1-$\sigma$
errorbars for each parameter do not include contributions from
correlations between the parameters, and may be affected by the
systematic residuals caused by the presence of substructure in the
host galaxy.

\begin{table}
\begin{center}	
\caption{The best-fitting two-dimensional PSF-convolved models to the
CL and LP images of the host galaxy of \object{GRB~980703}.}
\smallskip
\begin{tabular}{cccc}
\hline
\hline
 Parameter & $R^{1/n}$ & $R^{1/4}$ \\
\hline
CL Image & & \\
$R_e$     & $0\farcs117 \pm 0\farcs001$ & $0\farcs188 \pm 0\farcs001$ \\
${\langle\mu\rangle}_e$ & $20.38 \pm 0.01$            & $21.06 \pm 0.01$ \\
$\epsilon$              & $0.26 \pm 0.01$             & $0.26 \pm 0.01$ \\
$\theta_0$              & $24\fdg3 \pm 0\fdg2$        & $21\fdg9 \pm 0\fdg1$ \\
$n$                     & $1.07 \pm 0.01$             & 4 (fixed) \\
$\chi^2$/DOF            & 11.253                      & 13.050 \\
\hline
\hline
LP Image & & & \\
$R_e$      & $0\farcs134 \pm 0\farcs001$ & $0\farcs253 \pm 0\farcs001$ \\
${\langle\mu\rangle}_e$ & $20.27 \pm 0.01$            & $21.21 \pm 0.01$ \\
$\epsilon$              & $0.21 \pm 0.01$             & $0.17 \pm 0.01$ \\
$\theta_0$              & $10\fdg4 \pm 0\fdg5$        & $8\fdg6 \pm 0\fdg5$ \\
$n$                     & $1.02 \pm 0.01$             & 4 (fixed) \\
$\chi^2$/DOF            & 8.699                       & 9.879  \\
\hline
\hline
\end{tabular}
\label{TABLE:models}
\end{center}
\end{table}

     The standard de Vaucouleurs $R^{1/4}$ profile does not provide a
good fit to the host galaxy, whereas a generalized de Vaucouleurs
profile ($R^{1/n}$) with $n \approx 1$ provides a significantly better
fit.  However, both fits are formally strongly excluded by the data,
which is due to the fact that there are large residuals in the central
few pixels.  Fig.~\ref{FIGURE:host}b shows the CL image of the host
galaxy of \object{GRB~980703} with the best-fitting two-dimensional
PSF-convolved model subtracted.  The model was computed to a radius of
64 pixels ($= 1\farcs626 = 9.1 h_{100}^{-1}$ proper kpc) from the
centre of the galaxy.  The model provides a good fit for $R \ga
0\farcs25$, but there are systematic residuals in the inner $\approx
0\farcs25$ of the galaxy.  However, we note that the model assumes
that the galaxy has a smooth surface brightness distribution and does
not include small-scale structure, such as star-forming regions,
within the galaxy.  Therefore, we do not expect to obtain
$\chi^2/\mathrm{DOF} \approx 1$ for our fits.  The best-fitting model
provides a good fit over most of the host galaxy except for the
central few pixels, and the observed residuals are consistent with the
presence of substructure in the central regions of the galaxy.

     A Sersic~(\cite{S1968}) model with $n = 1$ corresponds to an
exponential disk.  Therefore, our modelling suggests that the host
galaxy of \object{GRB~980703} has a disk-like structure, with smaller
sub-structure in the central regions.  The excess light seen on the
west side of the galaxy after subtracting the best-fitting model (see
Fig.~\ref{FIGURE:host}b), may be a faint spiral arm.  The fitted
half-light radii of our Sersic~(\cite{S1968}) models correspond to
disk scale radii of $0\farcs196 \pm 0\farcs002$ on the CL image and
$0\farcs225 \pm 0\farcs002$ in LP.

     The derived half-light radius ($\approx 0.72 h_{100}^{-1}$ proper
kpc) is much smaller than the half-light radii seen in local late and
early type galaxies (Im et~al.~\cite{ICG1995}).  Phillips
et~al.~(\cite{PGG1997}) and Guzm{\'a}n et~al.~(\cite{GGK1997}) studied
galaxies in the Hubble Deep Field North that were \emph{selected} to
be compact (i.e., had measured half-light radii smaller than
$0\farcs5$ including the effect of the WFPC2 PSF). Their galaxies
covered a range of redshifts from $z \approx 0.4$ to $z
\approx 1.0$ and appear to have properties very similar to the host
galaxy of \object{GRB~980703}, namely $V\!-\!I$ colours in the range
$0.7 < V\!-\!I < 1.1$, half-light radii in the range $0.65
h_{100}^{-1} \le R_e \le 2.6 h_{100}^{-1}$ kpc (converted to our
cosmology and not corrected for the effect of the WFPC2 PSF), and
observed $I$-band magnitudes in the range $21 \le I \le 24$ for the
galaxies at redshifts similar to the redshift of \object{GRB~980703}.
The spectra of these galaxies are similar to the spectrum of the host
galaxy---i.e., dominated by strong emission lines from \ion{O}{ii},
and \ion{O}{iii}, as well as Balmer lines (Guzm{\'a}n
et~al.~\cite{GGK1997}; Djorgovski et~al.~\cite{DKB1998}).  The
rest-frame equivalent width (EW) of the
\ion{O}{ii} line from the host galaxy was found by Djorgovski
et~al.~(\cite{DKB1998}) to be 46 {\AA}, which is in the the middle of
the range $5 \le EW \le 94$ {\AA} found by Guzm{\'a}n
et~al.~(\cite{GGK1997}) for the compact galaxies in the Hubble Deep
Field North.

\subsection{The colour profile of the host galaxy}

\begin{figure}
\resizebox{\hsize}{!}{\includegraphics{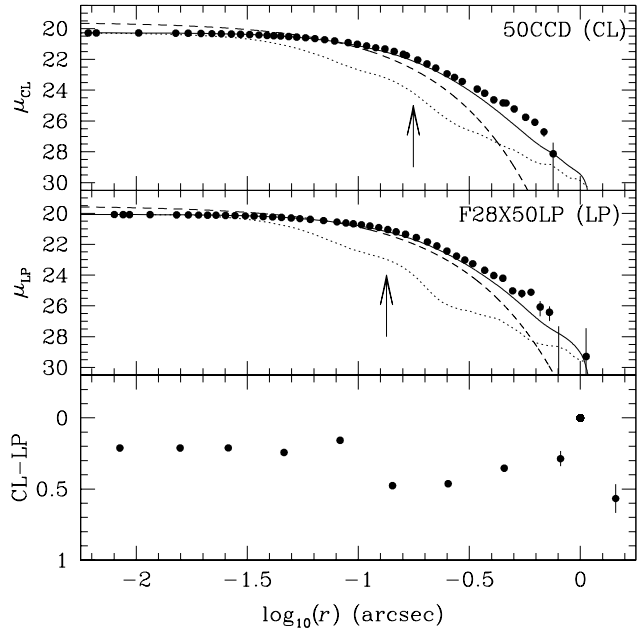}}
\caption{The top panel shows the azimuthally average surface
brightness profile that we derived from the \emph{HST}/STIS CL image
of \object{GRB~980703} (filled circles).  The solid line is the best
fitting PSF-convolved $r^{1/n}$ profile.  The dashed line shows the
best-fitting model after correcting for the PSF, and the dotted line
shows the point spread function for the image.  The arrow indicates
the half-light radius of the unconvolved model.  The middle panel
shows the same information for the LP band data.  The lower panel
shows the azimuthally averaged colour gradient in the host.  The data
has not been corrected for Galactic extinction or reddening.}
\label{FIGURE:profile}
\end{figure}

     The host galaxy of \object{GRB~980703} has $V_0 = 22.81 \pm 0.12$
and ${(V\!-\!R)}_0 = 0.40 \pm 0.15$ measured in an aperture with a
radius of $1\farcs524$ on the CAHA images.  This colour is consistent
with the colours predicted by Bloom et~al.~(\cite{BFK1998}) and
Vreeswijk et~al.~(\cite{VGO1999}) based on fitting a single power law
plus a constant flux to the optical decay of the OA\@.  We find no
evidence for azimuthal variations in colour, but the center of the
galaxy is $\approx 0.2$ mag bluer than the outer regions of the galaxy
with the step in the colour occurs at approximately the effective
radius of the best-fitting model (see Sect.~\ref{SECTION:morphology}).
Matthews \& Gallacher (\cite{MG1997}) note that this sort of trend in
colour is unique to late-type disc galaxies and some dwarf galaxies.
Fig.~\ref{FIGURE:profile} shows the surface brightness profiles and
the azimuthally averaged colour gradient that were derived from the
STIS CL and LP images.

\subsection{Star formation in the host galaxy\label{SECTION:sfr}}

     At a redshift of $z = 0.9661$ a rest-frame wavelength of 2800
\AA\ corresponds to the observer's $V$ band.  Therefore, we are able
to use Eq.~(2) of Madau et~al.~(\cite{MPD1998}) to estimate the star
formation rate in the host galaxy from the continuum flux at 2800
{\AA}.  For a Salpeter~(\cite{S1955}) initial mass function we find a
star-formation rate of $(4.1 \pm 0.5)h_{100}^{-2}$ $\mathcal{M}_{\sun}
\mathrm{yr}^{-1}$ while a Scalo~(\cite{S1986}) initial mass function
yields a star-formation rate of $(6.4 \pm 0.7)h_{100}^{-2}$
$\mathcal{M}_{\sun} \mathrm{yr}^{-1}$.  If we take $h_{100} = 0.7$ (de
Bernardis et~al.~\cite{dBA2000}) then the star-formation rate is
$\approx 8$--$13$ $\mathcal{M}_{\sun}$ yr$^{-1}$, depending on the
assumed initial mass function.  However, this calculation assumes that
there is no extinction in the host galaxy, which is probably an
incorrect assumption.  If $A_V = 0.3$ in the host then the
star-formation rate becomes $\approx 10$--$20$ $\mathcal{M}_{\sun}
\mathrm{yr}^{-1}$, while if if $A_V = 2.2$ in the host then the
star-formation rate is $\approx 65$--$100$ $\mathcal{M}_{\sun}
\mathrm{yr}^{-1}$.  These values are consistent with the
star-formation rate obtained by Djorgovski et~al.~(\cite{DKB1998})
using the strength of the [\ion{O}{ii}] emission line ($\approx 10$
$\mathcal{M}_{\sun} \mathrm{yr}^{-1}$ with $A_V = 0.3 \pm 0.3$).

\subsection{The luminosity of the host galaxy}

     Lilly et~al.~(\cite{LTM1995}) find a typical magnitude of
${(M^*_B)}_{AB} = -21.4$ for blue galaxies at $0.75 \le z < 1.00$, if
$(H_0,\Omega_m,\Omega_\Lambda) = (50,1,0)$.  For our adopted
cosmology, this corresponds to ${(M^*_B)}_\mathrm{AB} = -21.47 +
5\log_{10}(h_{100})$.  At $z = 0.9661$ the observed Kron--Cousins $I$
band is approximately half-way between the rest-frame Johnson $U$ and
$B$ bands.  If we assume that the host galaxy has a power-law spectrum
with $\beta_\mathrm{gal} = -1.08 \pm 0.59$ (from the CAHA data, see
Sect.~\ref{SECTION:obs_ground}) then $L_B \approx (3.2 \pm 0.7)
h_{100}^2 L^*_B$ where $L^*_B$ is the rest-frame $B$-band luminosity
of a typical blue galaxy at $z = 0.9961$.  Most of the uncertainty in
the luminosity comes from the uncertainty in the shape of the host
galaxy's rest-frame spectrum at $\lambda \la 4000$ {\AA}.  For
$h_{100} = 0.7$ the rest-frame $B$-band luminosity is $(1.6 \pm 0.4)
L^*_B$.  Therefore, the host is approximately an $L^*$ galaxy with a
total luminosity similar to that of other galaxies at $z = 0.9661$.
Our comparison of the total $B$-band luminosity of the host to $L^*_B$
is, however, somewhat uncertain because $M^*$ is highly correlated
with the slope of the faint end of the galaxy luminosity function, and
with its normalisation, $\phi^*$ (Lilly et~al.~\cite{LTM1995}).

     The specific star-formation rate per unit blue luminosity of the
host galaxy of \object{GRB~980703}, if we ignore extinction within the
host, is $\approx 6.5 \mathcal{M}_{\sun} \mathrm{yr}^{-1}
{L^*_B}^{-1}$, which is similar to that of other GRB host galaxies.
The specific star-formation rate is $\approx 0.6$ times that of the
hosts of \object{GRB~970508} ($11 \mathcal{M}_{\sun} \mathrm{yr}^{-1}
{L^*_B}^{-1}$) and \object{GRB~990123} ($11 \mathcal{M}_{\sun}
\mathrm{yr}^{-1} {L^*_B}^{-1}$) and $\approx 1.5$ times that of the
host of \object{GRB~990712} ($4.4 \mathcal{M}_{\sun} \mathrm{yr}^{-1}
{L^*_B}^{-1}$).  If the extinction in the host is non-negligible then
the true star-formation rate per unit blue luminosity may be much
higher.

     There is a faint ($\mathrm{CL} = 27.14 \pm 0.05$, $\mathrm{LP} =
26.29 \pm 0.06$) object $2\arcsec$ ($= 11.2 h_{100}^{-1}$ proper kpc)
south of the host galaxy (see Fig.~\ref{FIGURE:host}a), which may be a
companion galaxy. This object has FWHM $= 0\farcs097$, which is
slightly larger than the resolution of the CL image ($0\farcs084$) and
appears to have a faint extended structure around it.

\subsection{The position of GRB~980703 in the host galaxy\label{SECTION:astrometry}}

     The paucity of stars in the STIS images makes it difficult to
accurately determine the location of \object{GRB~980703} within the
host galaxy.  Only four stars are present in both the STIS CL image
and our ground-based images, and two of these are very faint in our
ground based images.  We used these four stars to determined the
location of the OA relative to the galaxy by comparing the STIS CL
image to an $R$-band image obtained with the NOT on 1998 July 7.163
(Pedersen et~al.~\cite{PKJ1998}).  At the epoch of the NOT
observation, the OA contributed $\approx 25$\% of the total host$+$OA
flux, so we registered, scaled, and smoothed the STIS image to match
the seeing and intensity of the NOT image, then subtracted the STIS
image from the NOT image.  The OA is found to be offset $0\farcs23$
west and $0\farcs02$ north of the center of the host galaxy with an
estimated 1$\sigma$ error of approximately $\pm 0\farcs10$.  This is
within approximately 1$\sigma$ of the location determined by Bloom
et~al.~(\cite{BKD2001}).  We find the GRB to have occurred $0\farcs23
\pm 0\farcs14$ from the centre of the host at a position angle of
$-85\fdg0 \pm 4\fdg6$, while Bloom et~al.~(\cite{BKD2001}) find that
the GRB occurred at a distance of $0\farcs11 \pm 0.09$ with a position
angle of $-28\fdg9 \pm 0\fdg3$.  The $1 \sigma$ error ellipses of
these two position determinations overlap, but we note that Bloom
et~al.~(\cite{BKD2001}) used deep Keck images to obtain an astrometric
solution involving 23 objects while our solution involved only four
stars.



\section{The Late-Time Light-Curve of the Optical Afterglow\label{SECTION:light}}

\begin{figure}
\resizebox{\hsize}{!}{\includegraphics{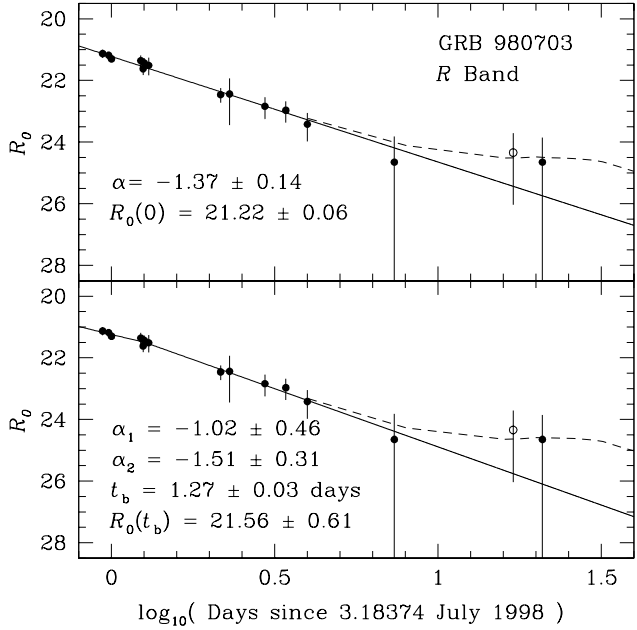}}
\caption{The upper panel shows the $R$-band photometry for the OA
associated with \object{GRB~980703} (filled circles) after subtracting
the light from the host galaxy ($R_\mathrm{gal} = 22.57 \pm 0.09$) and
correcting for extinction in the Milky Way ($A_R = 0.16 \pm 0.05$).
The open circle shows our NOT data point.  The solid line is the
best-fitting broken single law to all the data.  The dashed line is
the predicted $R$-band magnitude for a Type Ic SN like
\object{SN1998bw} at a redshift of $z = 0.9661$.  The lower panel
shows the same photometry with the best-fitting broken power law and
the corresponding prediction with an SN Ic like \object{SN1998bw}
included.}
\label{FIGURE:R_light_curve}
\end{figure}

\begin{figure}
\resizebox{\hsize}{!}{\includegraphics{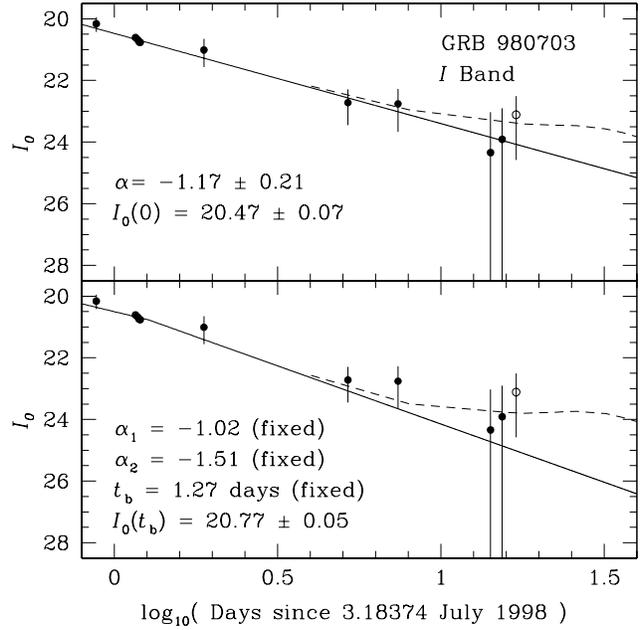}}
\caption{The upper panel shows the $I$-band photometry for the OA
associated with \object{GRB~980703} (filled circles) after subtracting
the light from the host galaxy ($I_\mathrm{gal} = 22.02 \pm 0.23$) and
correcting for extinction in the Milky Way ($A_I = 0.11 \pm 0.04$).
The open circle shows our NOT data point.  The solid line is the
best-fitting single power law to all the data.  The dashed line is the
predicted $I$-band magnitude for a Type Ic SN like \object{SN1998bw}
at a redshift of $z = 0.9661$.  The lower panel shows the same
photometry with the best-fitting broken power law and the
corresponding prediction for an SN Ic like \object{SN1998bw}.  These
data should be considered less reliable than the $R$-band data since
$I_\mathrm{gal}$ was not measured directly, but determined by assuming
that the host has a power-law spectrum with $\beta_\mathrm{gal} =
-1.08 \pm 0.59$ (see Sect.~\ref{SECTION:obs_ground}).}
\label{FIGURE:I_light_curve}
\end{figure}

     Figs.~\ref{FIGURE:R_light_curve}~and~\ref{FIGURE:I_light_curve}
shows the $R$- and $I$-band light curves for the OA based on
measurements taken from the literature and on our own NOT
measurements. The magnitudes have been corrected for the contribution
from the underlying host galaxy and for Galactic extinction.  The
light from the host galaxy was removed by subtracting the total light
from the host galaxy as determined from our CAHA images.

     We fit both a single power law ($f_{\nu}(t) = f_{\nu}(0)
t^\alpha$), where $f_{\nu}(t)$ is the flux in $\mu$Jy $t$ days after
the burst, and a broken power law

\begin{equation}
f_{\nu}(t) = \left \{ 
        \begin{array}{lll} 
                f_{\nu}(t_b) {(t/t_b)}^{\alpha_1}, & 
                \mathrm{if}  &  t \le t_b \\ 
                f_{\nu}(t_b) {(t/t_b)}^{\alpha_2}, & 
                \mathrm{if}  &  t > t_b, 
        \end{array} 
             \right. 
\label{EQUATION:broken_power_law} 
\end{equation} 

\noindent
where $\alpha_1$ is the slope before the break, $\alpha_2$ is the
slope after the break, and $t_b$ is the time in days after the burst
that the break occurred, to the $R$-band data as described in Holland
et~al.~(\cite{HBH2000}) and Jensen et~al.~(\cite{JFG2001}).  Only data
$t \le 10$ days after the burst were used for the fit since a Type Ic
supernova (SN) like \object{SN1998bw} would not make a significant
contribution to the total flux at these times.  The parameters of our
best fits are listed in Table~\ref{TABLE:light_curve_fits} and the
fits themselves are shown in
Figs.~\ref{FIGURE:R_light_curve}~and~\ref{FIGURE:I_light_curve}.  The
small $\chi^2/\mathrm{DOF}$ values for both the single and broken
power law fits reflect the large formal uncertainties in the late time
data that arise from subtracting the OA from the host at a time when
both the OA and the host had similar magnitudes.  The small number of
data points makes it impossible to fit a broken power law in the
$I$-band, so we scaled the broken power law that was fit to the
$R$-band data to match (in a chi-square sense) the early $I$-band
photometry.  The best fit occurred for $I_0(t_b) = 20.77 \pm
0.05$. The broken power law is formally a slightly better fit to the
$R$-band light curve than a single power law is, although an F-test
suggests that the broken power law is only significant at about the $1
\sigma$ level.  There are several gaps in the decay curve,
and large error bars in the late-time data due to the subtracting of
the light from the bright host galaxy, so we cannot unambiguously
conclude that there is a break in the light curve.  Therefore, we
prefer the single power law fit since it requires fewer free
parameters to fit the light curve.  We note, however, that breaks have
been seen in most well-sampled GRB light curves, and that the time and
magnitude of the break for \object{GRB~980703} is consistent with the
breaks seen for several other OAs.

\begin{table}
\begin{center}	
\caption{The parameters of the best-fitting single power law and
broken power law to the decay of the optical light curve of
\object{GRB~980703}.}
\smallskip
\begin{tabular}{cccc}
\hline
\hline
                    & $R$               & $I$ \\
\hline
Single Power Law    &                   & \\
 $\alpha$           & $-1.37 \pm 0.14$  & $-1.17 \pm 0.21$ \\
 $f_0$ ($\mu$Jy)    & $9.805\pm 0.512$  & $15.373 \pm 0.930$ \\
 mag$_0$(0)         & $21.22 \pm 0.06$  & $20.47 \pm 0.07$ \\
 $\chi^2$/DOF       & 0.268             & 0.322 \\
 DOF                & 11                & 6 \\
\hline
Broken Power Law    &                   & \\
 $\alpha_1$         & $-1.02 \pm 0.46$  & $-1.02$ (fixed) \\
 $\alpha_2$         & $-1.51 \pm 0.31$  & $-1.51$ (fixed) \\
 $t_b$ (days)       & $1.27 \pm 0.03$   & $1.27$ (fixed) \\
 $f(t_b)$ ($\mu$Jy) & $7.541 \pm 0.061$ & $11.750 \pm 0.584$ \\
 mag$_0$($t_b$)     & $21.51 \pm 0.61$  & $20.77 \pm 0.05$ \\
 $\chi^2$/DOF       & 0.231             & 0.496 \\
 DOF                & 9                 & 7 \\
\hline
\hline
\end{tabular}
\label{TABLE:light_curve_fits}
\end{center}
\end{table}

     There is growing evidence that at least some GRBs are associated
with supernovae.  The prototype for this association is the temporal
and positional coincidence of \object{GRB~980425} and the unusual Type
Ib/c \object{SN1998bw} (Galama et~al.~\cite{GVP1998}). VLT and
\emph{HST}/STIS images show that this GRB/SN occurred in a
star-forming region in the sub-luminous barred spiral galaxy
\object{ESO~184$-$G82} (Sollerman et~al.~\cite{SKF2000}; Fynbo
et~al.~\cite{FHA2000}). Evidence for a GRB/SN association has also
been seen for \object{GRB~980326} (Castro-Tirado \&
Gorosabel~\cite{CTG1999}; Bloom et~al~\cite{BKD1999}),
\object{GRB~970228} (Dar~\cite{D1999}; Reichart~\cite{R1999}; Galama
et~al.~\cite{GTV2000}), \object{GRB~970514} (Germany
et~al.~\cite{GRS2000}), \object{GRB~990712} (Fruchter
et~al.~\cite{FVH2000}; Bj{\"o}rnsson et~al.~\cite{B2001}), and
\object{GRB~991002} (Terlevich et~al.~\cite{TFT1999}).  In addition,
Hudec et~al.~(\cite{HHH1999}) have identified five more candidates for
SN/GRB associations based on positional and temporal coincidence.  In
Figs.~\ref{FIGURE:R_light_curve}~and~\ref{FIGURE:I_light_curve} we
show the predicted light curves for \object{GRB~980703} if it was
associated with a Type Ic SN like \object{SN1998bw}.  Although both
the $R$- and $I$-band data are consistent with both a pure power law
decline and a broken power law decline the addition of an SN Ic like
\object{SN1998bw} results in a better fit to the data. However, the
large error bars in the late-time data prevent us from being able to
reliably distinguish between a SN and a no-SN scenario for
\object{GRB~980703}.

     It has recently been suggested that a dust echo could produce an
excess flux in the decay of the late-time optical light curve of a GRB
similar to the flux excesses seen in \object{GRB~970228} and
\object{GRB~980326} (Esin \& Blandford~\cite{EB2000}).  Using their
method we predict an excess $R$-band flux due to dust 17 days after
the burst of $F^E_{\nu_\mathrm{ob}} \approx 0.03$ $\mu$Jy,
approximately $2 \sigma$ less than the observed excess flux at this
time, assuming a single power law decay.  Therefore, we are unable to
distinguish between the dust echo and no dust echo cases.

     The $R$-band slope of the best-fitting single power law is
$\alpha = -1.37 \pm 0.14$, which is consistent with other values in
the literature.  Combining the optical slope with the observed slope
of the $X$-ray decay ($\alpha_X < -0.91$, Vreeswijk
et~al.~\cite{VGO1999}), and assuming that the GRB is a collimated
outflow into a region with a $\rho(r) \propto r^{-\delta}$ density
profile (e.g.~Panaitescu et~al.~\cite{PMR1998}, M{\'e}sz{\'a}ros
et~al.~\cite{MRW1998}), implies that $\delta = 2$ (Panaitescu \&
Kumar~\cite{PK2001}).  This suggests that the fireball is expanding
into a local medium dominated by a pre-existing stellar wind, similar
to the situation that Jaunsen et~al.~(\cite{JHB2001}) found for
\object{GRB~980519}.  The optical spectrum of the early OA (1.2 days
after the burst) has a slope of $\beta_\mathrm{OA} = -2.71 \pm 0.12$
(Vreeswijk et~al.~\cite{VGO1999}).  For an expansion into a
pre-existing stellar wind the spectral index of the OA is given by
$\beta_\mathrm{OA} = (1 + 2\alpha) / 3 = -0.58 \pm 0.09$, which is
inconsistent with the observed early-time slope of OA's optical
spectrum.  This discrepancy, however, can be resolved if there are
several magnitudes of extinction along the line of sight to the GRB\@.
The extinction measurements in Sect.~\ref{SECTION:intro} are
consistent with $A_V$ decreasing by $\approx 0.3$ mag/day from $A_V
\approx 1.5$ one day after the burst to $A_V \approx 0.5$ five days
after the burst.  An examination of the spectrum of Djorgovski
et~al.~(\cite{DKG1998}) suggests that $\beta_\mathrm{OA} \approx -1.6
\pm 0.2$ in the $R$ band 4.4 days after the burst, which suggests that
the burst evolved to become bluer at later times, which is consistent
with $A_V$ decreasing with time.  This suggests that some of the
discrepancy between the observed spectral index and the spectral index
which is predicted from the decay of the optical and $X$-ray light
curves may be due to the time evolution of the OA's spectrum that
result from changes in the amount of dust along the line of sight to
the burst during the first few days after the burst.


\section{Conclusions\label{SECTION:conc}}

     The host galaxy of \object{GRB~980703} is a compact blue,
star-forming galaxy with $V_\mathrm{gal} = 23.00 \pm 0.10$ and
${(V\!-\!R)}_\mathrm{gal} = 0.43 \pm 0.13$.  The central regions of
the host are $\approx 0.2$ mag bluer than the outer regions.  Its
rest-frame $B$-band luminosity is $L_B \approx 3.2 h_{100}^2 L^*_B$,
and it has a star-formation rate, based on the integrated flux at 2800
{\AA}, of $8$--$13$ $\mathcal{M}_{\sun} \mathrm{yr}^{-1}$, assuming no
extinction in the host.  This suggests that the host galaxy is
undergoing a phase of active star formation similar to what has been
seen in other GRB host galaxies.  We find that the galaxy is best fit
by a Sersic~(\cite{S1968}) $R^{1/n}$ profile with $n \approx 1$ and a
half-light radius of $0\farcs13$ ($= 0.72 h_{100}^{-1}$ proper kpc).
This corresponds to an exponential disk with a scale radius of
$0\farcs22$ ($= 1.21 h_{100}^{-1}$ proper kpc).  The residuals of the
fit show that the morphology of the host is irregular, but we are
unable to determine if the location of \object{GRB~980703} coincided
with any special features in the host.  The galaxy resembles the
compact star-forming galaxies that Phillips et~al.~(\cite{PGG1997})
and Guzm{\'a}n et~al.~(\cite{GGK1997}) studied galaxies in the Hubble
Deep Field North.

     The optical decay of \object{GRB~980703} followed the same broad
pattern established by other long-duration GRBs: a power-law decay,
perhaps with a break a few days after the burst, possibly followed by
an increase in flux $\approx 10(1+z)$ days after the burst.


\begin{acknowledgements}

     This work was supported by the Danish Natural Science Research
Council (SNF).  JG acknowledges the receipt of a Marie Curie Research
Grant from the European Commission.  MA acknowledges the Astrophysics
group of the Physics Department of University of Oulu for support of
his work.  GB acknowledges partial support from the Icelandic Research
Council and the University of Iceland Research Fund.  We would also
like to thank the anonymous referee for a careful reading of the
original manuscript, and for useful comments that greatly improved
this paper.

\end{acknowledgements}


  
\end{document}